\documentclass[
a4,12pt
  ]
{article}

\usepackage{amsmath,amssymb,graphicx,url}
\usepackage[affil-it]{authblk}

\begin{document}

\newcommand{\classification}[1]{}
\newcommand{\keywords}[1]{}
\newenvironment{theacknowledgments}
{\vspace{5mm}\centerline{\textbf{Acknowledgments}}
\vspace{5mm}
\noindent\ignorespaces}%
{\par\noindent%
\ignorespacesafterend}

\title{Data analysis challenges in transient gravitational-wave astronomy}

\classification{95.85.Sz, 04.80.Nn, 95.55.Ym}

\author{\'Eric Chassande-Mottin for the LIGO Scientific Collaboration and the Virgo Collaboration}

\affil{APC, Univ Paris Diderot, CNRS/IN2P3, CEA/Irfu, Obs de Paris, Sorbonne Paris Cit\'e, France}

\maketitle

\begin{abstract}
  Gravitational waves are radiative solutions of space-time dynamics predicted
  by Einstein's theory of General Relativity. A world-wide array of large-scale
  and highly sensitive interferometric detectors constantly scrutinizes the
  geometry of the local space-time with the hope to detect deviations that would
  signal an impinging gravitational wave from a remote astrophysical
  source. Finding the rare and weak signature of gravitational waves buried in
  non-stationary and non-Gaussian instrument noise is a particularly challenging
  problem. We will give an overview of the data analysis techniques and
  associated observational results obtained so far by Virgo (in Europe) and LIGO
  (in the US), along with the prospects offered by the up-coming advanced
  versions of those detectors.
\end{abstract}

Einstein's theory of General Relativity introduces the concept of a deformable
and evolving space-time. The dynamics of space-time is prescribed by the
Einstein equation. In linearized gravity which assumes small deformations
in a nearly flat space-time, this equation reduces to the wave equation which
therefore evidences the existence of radiative solutions. The latter are
referred to as \textit{gravitational waves} (GW) and can be phenomenologically seen
as propagating disturbances of space-time itself. The theory also predicts that
GW are transverse waves, that they nominally propagate at the speed of light and
possess two independent polarizations
\cite{thorne87:_gravit_radiat,maggiore08:_gravit_waves_volum}.

GW have never been directly detected, i.e. through the measurement of their
effect on a man-made instrument. Strong evidence of their existence has been
provided by the observation of the famous Hulse-Taylor pulsar binary
(PSR~B1913+16) \cite{weisberg10:_timin_measur_relat_binar_pulsar}. The decay rate
of the binary orbital period is in remarkable agreement with the predicted
evolution obtained under the assumption that this system radiates energy away
in the form of GW.

The direct search for GW made notable progress with the advent of dedicated
instruments based on high-precision laser interferometry such as LIGO and Virgo
(see \cite{sathyaprakash09:_physic_astrop_cosmol_gravit_waves, riles12:_gravit}
for a detailed review). With the ongoing installation of a new and improved
generation of those instruments, the first discovery is expected within the decade.

While electromagnetic waves are produced by accelerated charges, GW are produced
by accelerated masses.  Very large masses and relativistic velocities are
necessary to generate GW at a detectable level. For this reason, the current
projects aiming at detecting GW target potential astrophysical sources involving
very dense and compact objects such as neutron stars or black holes. Very
energetic astrophysical events such as the coalescence of neutron star and/or
black hole binaries, or stellar core collapses are expected to be the source of
intense and short-duration bursts of GW
\cite{sathyaprakash09:_physic_astrop_cosmol_gravit_waves}.

Because of the limited rate of occurrence of such events, searching for such
transient GW in the LIGO and Virgo data essentially consists in searching for
rare and weak signals at the detectability limit. This article reports the
state-of-the-art of the search for GW transient signals with a focus on the
related data analysis challenges. Searches for long-lived signals such as
periodic GWs from rotating neutron stars and stochastic GW backgrounds are
beyond the scope of this paper. We first give some introductory material
with a general presentation of the detectors in Sec.~\ref{sec:detectors} and a
review of the relevant astrophysical sources in
Sec.~\ref{sec:sources}. Sec.~\ref{sec:DA} gives an overview of the major problems
faced when searching for transient GW along with the data analysis methods
deployed to address them.

\section{Interferometric GW detectors}
\label{sec:detectors}

\begin{figure}
  \centerline{\includegraphics[width=\textwidth]{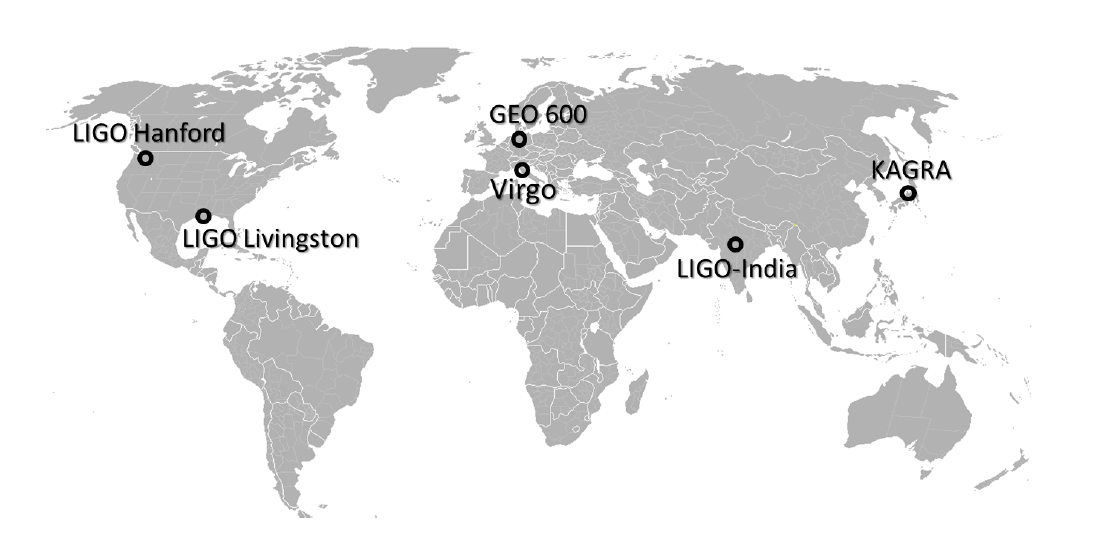}}
  \caption{\textbf{Geographic location of the current and future GW
      interferometric detectors}. This world map displays the location of the
    four sites of the first generation detectors (LIGO H and L, Virgo and GEO),
    and six sites of the second generation (complemented by LIGO I and KAGRA).
    The future detector LIGO India is still pending approval and its exact
    location is yet to be determined. Credits: \cite{shawhan12:_rapid}}
  \label{fig:map}
\end{figure}

The first generation of interferometric GW detectors comprises five large-scale
instruments in total (see Fig.~\ref{fig:map}). The US-based Laser Interferometer
Gravitational-Wave Observatory (LIGO)~\cite{abbott09:_ligo} includes three
kilometric-scale instruments located in Livingston, Louisiana (labelled L1) and
Hanford, Washington (the latter hosting two interferometers in the same vacuum
enclosure with labels H1 and H2). The French-Italian project
Virgo~\cite{accadia12:_virgo} has one instrument of the same class located in
Cascina near Pisa, Italy (labelled V1). This set of kilometer-scale instruments
is complemented by a detector with more modest dimensions (several hundreds of
meters): GEO~\cite{grote10:_geo} (labelled G1), a German-British detector in
operation near Hanover, Germany.

\begin{figure}
  \centerline{
  \begin{tabular}{c}
  \includegraphics[width=\textwidth]{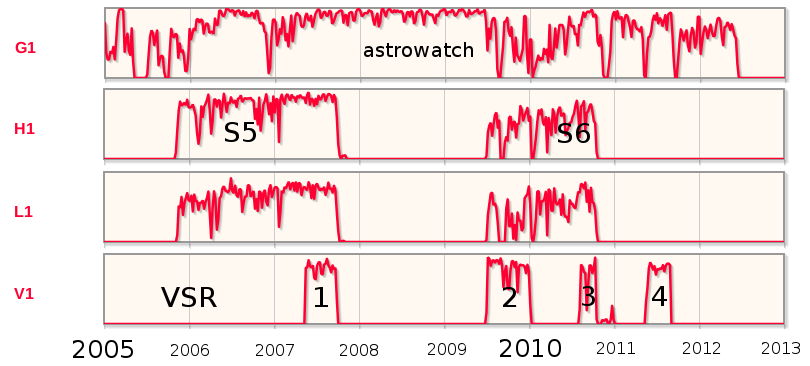}\\
  \includegraphics[width=\textwidth]{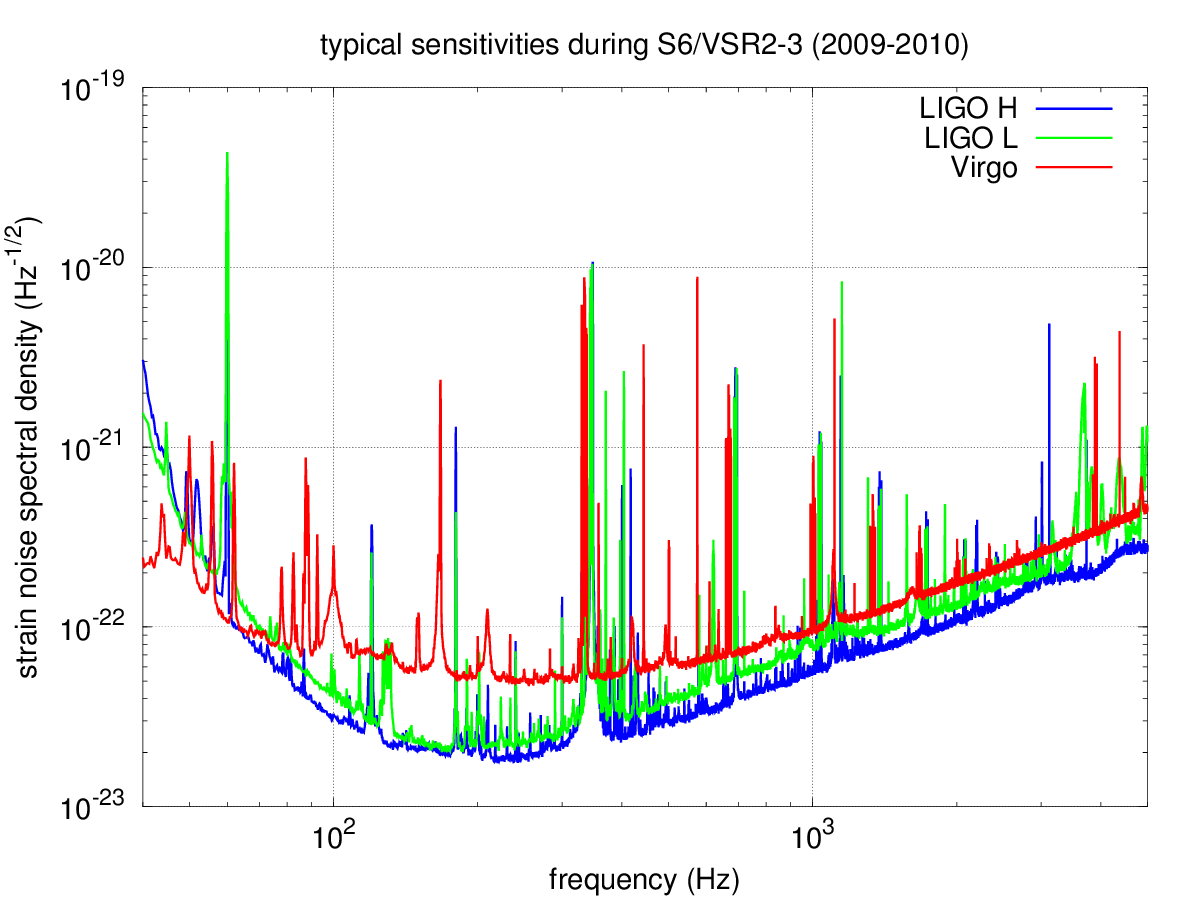}
  \end{tabular}}
  \caption{\textbf{(top) Time line of the data takings completed so far.}
    Credits: \cite{ligo_open_scien_center} \textbf{(bottom) Sensitivity achieved
      by LIGO and Virgo during their last science data taking S6/VSR2--3} \cite{abadie12:_sensit_achiev_ligo_virgo_gravit}.}
  \label{fig:sensitivity}
\end{figure}

Despite major differences in the technologies in use, all those instruments
measure gravitational waves through the same principle. They all sense the
strain that a passing GW exerts on space-time by monitoring the differential
length $\delta \ell$ of the optical path followed by two laser beams propagating
along orthogonal directions over a distance $L$. This is performed by letting
the two beams interfere similarly to the Michelson-Morley experiment. The
interference is closely related to the difference in the phase accumulated by
the two beams before they combine and hence to the difference in their optical
paths. The measurement of the interference light power allows that of $\delta
\ell$ with high accuracy.  Measurement noises (mainly the thermal noise due to
the Brownian agitation of the atoms constitutive of the optics and the shot
noise due to the quantum nature of light) can be reduced to reach the level of
$h \equiv \delta \ell/L \sim 10^{-21}$, where the detector response $h$ is
directly connected to the dimensionless amplitude $h_+$ and $h_\times$ of the
two GW polarizations\footnote{These quantities measure the strain or fractional length change that a GW exerts on space-time and are therefore dimensionless}. The best sensitivity is achieved in a frequency band
ranging from $\sim 100$ Hz to $1$ kHz approximately (see
Fig.~\ref{fig:sensitivity} -- bottom).

The detector response is a linear mixture $h=F_+ h_+ + F_\times h_\times$ of the
two GW polarizations. The antenna pattern factors $F_+$ and $F_\times$
characterize the way the wave polarizations couple to the detector.  The
coupling ${\cal F}=(F_+^2+F_\times^2)^{1/2} \leq 1$ is maximum for waves
impinging perpendicularly to the detector plane and is minimum (and exactly
zero) for waves from the four ``blind'' directions associated to the two
bisectors of the detector arms. GW detectors are non-directional instruments as
${\cal F} \gtrsim 1/2$ for more than half of the sky.

The first generation detectors have conducted a series of science data takings
reaching an integrated observation time of about 2 years (see Fig.~\ref{fig:sensitivity}
-- top). The data takings are coordinated in order to maximize the observation
time with the three most sensitive detectors operating while always maintaining
at least one detector in ``astro-watch'' mode in case of an outstanding galactic
event.

\begin{figure}
 \centerline{\includegraphics[width=\textwidth]{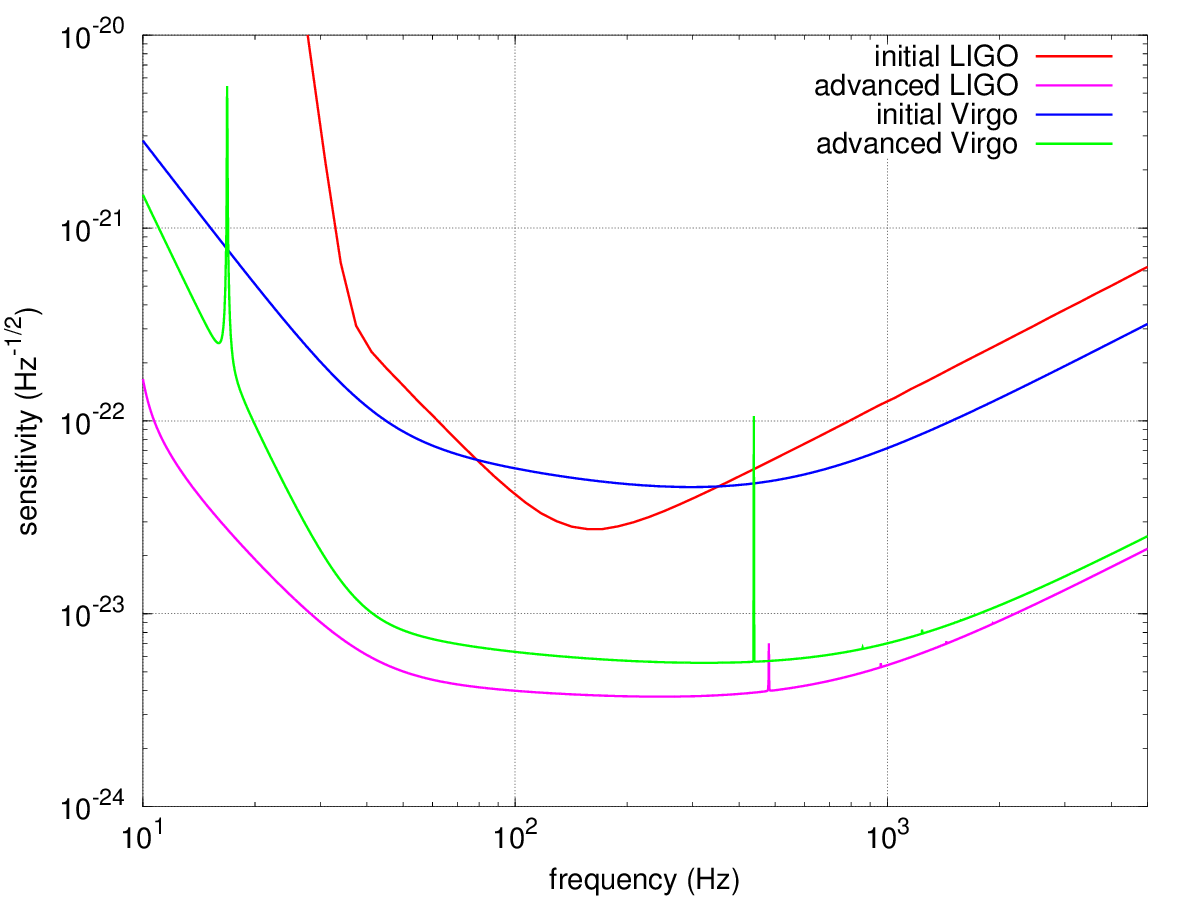}}
  \caption{\textbf{Projected sensitivities for the advanced LIGO and advanced
      Virgo detectors compared with the design sensitivity of their initial
      version}.}
  \label{fig:advancedsensitivities}
\end{figure}

The first generation of detectors has now been decommissioned and it is
currently being replaced by a second generation of ``advanced''
detectors. Thanks to major upgrades in their infrastructure and instrumentation,
a ten-fold increase in sensitivity is expected with the advanced detectors as
indicated in Fig.~\ref{fig:advancedsensitivities}. The GW amplitude decaying
inversely with the distance, this corresponds to a factor of thousand in the
observable volume and hence in the number of detectable sources.
Advanced detectors are likely to detect several tens and possible several
hundreds of sources as we will see in the next Section.

The installation of the advanced LIGO detectors \cite{harry10:_advan_ligo}
should be completed by the end of 2013 and a first science run is likely to
take place in 2015. The original plan was to install two four-kilometer
detectors at Hanford site, but there is now a proposal (still to be approved by
U.S. and Indian institutions) to move one of those detectors to a new
observatory in India. If this plan materializes, the third detector at the
Indian site would start operation around 2020.  Advanced Virgo plans to have a
robustly operating detector in 2015 and to begin collecting science data as soon
as possible after that \cite{accadia12:_advan_virgo_techn_desig_repor}. GEO
foresees a program of upgrades ``GEO-HF'' \cite{lueck10:_geo} which focuses on
improving the detector sensitivity at high-frequencies thanks to a larger laser
power and the use of ``squeezed light''. The network of advanced detectors will
be completed by the Japanese KAGRA detector \cite{somiya12:_detec_kagra_japan}
which has the specificity of being installed underground in the Kamioka mine
(where the seismic motion is much lower than at the surface) and to operate at
cryogenic temperatures to reduce thermal noise. An initial three-kilometer
room-temperature interferometer is expected to be operational by 2015, with the
full cryogenic interferometer ready to start taking data by 2018.

\section{Astrophysical sources of GW transients}
\label{sec:sources}

Phenomenologically, GW emission arises from relativistic bulk motion.  At lowest
order, GW can be related to variations in the quadrupolar moment of the mass
distribution \cite{thorne87:_gravit_radiat}. Therefore, GW sources have to
present some degree of non-axisymmetry. In this section, we review the
astrophysical scenarios giving rise to GW emission. The coalescence of
neutron-star and/or black-hole binaries similar to the Hulse-Taylor binary
mentioned previously is often considered the most promising one.

The last minutes before the system merges give rise to the emission of an
intense burst of GW. Post-Newtonian expansions of the binary dynamics
\cite{blanchet02:_gravit_newton,buonanno07:_inspir} are used to predict the
gravitational waveforms radiated during the inspiral phase which precedes the
merger. The GW signature consists in a \textit{chirp} signal whose frequency
sweeps towards high values according to a power law at first order 
. A substantial amount of energy is radiated
in the following phase when the two bodies merge into a black hole. In this
highly relativistic phase, the perturbative treatment of binary dynamics is not
valid anymore and one has to resort to numerical simulations. The process is
concluded by the ring-down phase during which the resulting distorted black hole
radiates away its asymmetry down to equilibrium. During the whole coalescence
process, a stellar-mass binary with equal masses radiates away of order of a
percent of its rest
mass\cite{sathyaprakash09:_physic_astrop_cosmol_gravit_waves}.

Although binary systems are fairly common, only a small fraction eventually
forms a compact binary that is sufficiently tight to coalesce in less than
Hubble time. A survey of population estimates \cite{abadie10:_predic} gives a
``realistic'' rate of one neutron-star--neutron-star
coalescence\footnote{Similar rates are obtained for the other types of systems
  mixing neutron stars and/or black-holes.} per 10,000 years per galaxy
equivalent in size to the Milky Way. GW detectors can ideally observe those
binary systems up to a distance of $\sim$ 30 Mpc and $\sim 440$ Mpc for initial
and advanced detectors resp. \cite{abadie10:_predic}. Converted into a rate of
\textit{detectable} coalescences, this leads to $\sim 0.02$ events per year with
the first generation of (initial) detectors and to $\sim 40$ events for the
second generation (advanced).  Large error bars are attached to those estimates
reflecting the weakness of the observational constraints we have about those
systems. The above stated rates can then be 10 times smaller or larger in the
``pessimistic'' or ``optimistic'' scenarios respectively.  The ``realistic''
rates presented above are corroborated by the ones derived assuming that compact
binary mergers are the progenitors of short-hard gamma-ray bursts (GRB)
\cite{abadie10:_predic}.


Gravitational stellar-core collapse is another potential source of GW if some
degree of non-axisymmetry is exhibited during this process. The simulations
required to make reliable predictions of the emission levels are very
challenging as they have to incorporate many physical ingredients including
relativistic magneto-hydrodynamics and a detailed treatment of neutrino
transport and nuclear interactions
\cite{ott09:_gravit_wave_signat_core_collap_super}. The current realistic
estimate of the amount of radiated GW energy is of order $10^{-7} M_\odot$ and
corresponds approximately \cite{sutton:_rule_thumb_detec_gravit_wave_burst} to a
distance reach of order $\sim 10$ kpc with the initial detectors, $\sim 100$ kpc
with the advanced detectors. The detectable sources are therefore located in the
Galaxy.

Another potential source of GW bursts are ``neutron-star quakes''
\cite{chamel08:_physic_neutr_star_crust} during which the vibrational normal
modes of a neutron star are excited and damped by GW emission. Star quakes may
origin from the disruption of the star crust due to the sudden rearrangement of
the magnetic field of a highly-magnetized neutron star (magnetar). Cosmic string
cusps may be also listed among the potential GW burst sources
\cite{siemens06:_gravit}.

\section{Searches for GW transient signals}
\label{sec:DA}

\subsection{Time-series analysis}
\label{sec:timeseries}

In detection problems, the availability of \textit{a priori} information plays a
major r\^ole.  We have seen in Sec.~\ref{sec:sources} that the GW signature from
coalescing binaries of neutron stars and/or black holes have a specific time
evolution which can be predicted with good accuracy. This morphological
information helps to distinguish a real GW signal from the instrumental or
environmental noise. The search for known signals is efficiently performed by
\textit{matched filtering techniques} \cite{schutz91:_gravit_waves} which
cross-correlates the data with the expected ``template'' waveforms obtained from
the source model.

Because of the highly-relativistic dynamics associated with the production of GW,
some of the expected GW waveforms are difficult to predict with accuracy. This
calls for detection methods that are robust to the model
uncertainties. \textit{Excess power methods} essentially consist in searching
for a broad family of GW waveforms by scanning a time-frequency map for
transient excursions. The time-frequency map is obtained by projecting the data
onto a dictionary of elementary waveforms that tiles the time-frequency
plane. Several types of dictionary have been tested including local cosines
\cite{anderson01:_exces}, sine-Gaussian wavelets \cite{chatterji04:_multir},
orthogonal wavelet packet bases \cite{klimenko04:_perfor_waveb_ligo} or
chirplets \cite{chassande-mottin10:_detec}. Real GW signals are unlikely to
correlate exactly with one element in the dictionary, but with several of
them. Clustering algorithms are generally applied to harvest the signal energy
scattered over several elements
\cite{sylvestre02:_time,khan09:_enhan_ligo,chassande-mottin06:_best}.

The time-frequency dictionaries mentioned above are composed of ``generic''
elementary waveforms mainly motivated by mathematical or algorithmic
arguments. Astrophysically motivated dictionaries can be obtained by extracting
the relevant information from catalogs of GW signals developed through numerical
simulations \cite{heng09:_super,rover09:_bayes,brady04:_incor,chassande-mottin03:_learn}.

\subsection{Multi-detector analysis}
\label{sec:multidetector}

We described the basic ideas employed to analyze the data stream from individual
detectors. A gain in sensitivity is expected from the availability of a joint
observation by multiple detectors. This section discusses several aspects
related to the combined analysis of multiple detector data.

\subsubsection{Coherent analysis of multiple data streams}
\label{sec:coherent}

We already mentioned that the detector receives a mixture of both GW
polarizations which depends on the relative orientation and alignment of the
detector and wave. Since the detectors are not co-planar and co-aligned, they
couple differently to the incoming wave resulting in observed responses with
different initial phases and amplitudes. Because of its finite speed, a GW reaches
the detectors at different times. All those differences can be exploited using
\textit{coherent analysis techniques} to improve the overall sensitivity.  Those
techniques consist in compensating the phase shift and delay of the various
responses to align them in time and phase assuming a given
direction-of-arrival. The resulting data streams are combined so that the sum
operates constructively for GW signals from the selected direction. The data
stream which results from the coherent combination maximizes the signal-to-noise
ratio (SNR). The combined stream can then be analyzed using methods inspired
by the single detector case, i.e., excess power methods for the unmodelled GW
bursts \cite{guersel89:_near, klimenko08:_coher} and matched filtering
techniques \cite{harry11,pai01:_data} for inspiralling binaries. The coherent
analysis being directional (each coherently combined stream is associated to a
given direction), the outcome is a probability (pseudo-)distribution over the
sky usually referred to \textit{sky map} from which the most likely location of
the source can be extracted.

\subsubsection{Mitigation of non-Gaussian/non-stationary noise}
\label{sec:glitch}

The noise of the real instruments is far from the ideal properties of
stationarity and Gaussianity we expect from the main fundamental (thermal and
quantum) noises. The tails of the noise distribution is dominated by
a non-Gaussian and non-stationary component consisting in a
large number of transient noise excursions commonly called
\textit{glitches}. Glitches are produced by a variety of environmental and
instrumental processes, such as upconversion of seismic noise or saturations in feedback
control systems. Since glitches occasionally occur nearly simultaneously in
separate detectors by chance, they can mimic a gravitational-wave signal.

The population of glitches is difficult to model. The size and the large
complexity of GW detectors makes this modelling even more difficult. GW
detectors being instruments extended over kilometers, it is hard to completely
isolate them from the outside world and the surrounding anthropic
activity. Therefore, the accurate modelling of the non-Gaussian/non-stationary
noise background is for now out-of-reach. It has to be mitigated and this can
be done at least partially by using multiple data.

It is possible to calculate combinations of the data from multiple detectors
where the GW signals from all detectors interfere destructively in the sum. The
GW signal thus cancels, but not background glitches. The energy in these
``null'' stream(s) may be used to reject or down-weight events not consistent
with a gravitational wave \cite{chatterji06:_coher,klimenko08:_coher}. The
success of such tests depend critically on having several independent detectors
of comparable sensitivity.

\subsubsection{Use of multiple detector data for background estimation}
\label{sec:timeshift}

We explained earlier that the accurate modelling of non-Gaussian non-stationary
noise is out-of-reach. The remaining part of the glitches that cannot be
identified by the coherent techniques described in the previous section
constitutes the dominating background noise in burst searches. This background
has to be estimated. However, GW signals cannot be turned off: the detectors
cannot be shielded from them. Therefore, we don't have ``noise-only'' data at
our disposal for background estimation.

Nevertheless, the background can be estimated thanks to the availability of
multiple data streams by time shifting one detector's data. The time shift is
chosen to be much longer than the time-of-flight between detectors ($\sim 30$
ms) and coherence time scale of the detector noise ($\sim$
seconds). The time-shifted (or ``time-slide'') analysis leaves only triggers due to
accidental coincidences of instrumental glitches. The contribution from real GW
signals is practically erased. By repeating the analysis many times with different
time shifts, we get an accurate estimate of the rate of background events. For
sufficiently large time shifts, each trial can be considered independent of the
other. However, the number of time slides cannot be increased indefinitely as a
significant correlation between time slides will occur above a certain level
\cite{was10}.

The p-value measuring the significance of a GW event can be computed by
computing the fraction of louder background events from the time-slide analysis.

\subsection{Data quality}
\label{sec:DQ}

\begin{figure}
  \centerline{\includegraphics[width=\textwidth]{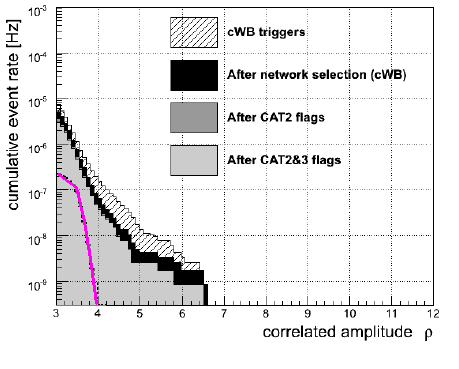}}
  \caption{\textbf{Examples of background distributions for the coherent
      Wave-Burst algorithm \cite{klimenko08:_coher} for a three-detector network
      (LIGO-H and L and Virgo) during S6-VSR2/3 data taking.} This distribution
    is given as a function of the correlated amplitude $\rho$ homogeneous to the
    signal-to-noise ratio. \textbf{Hatch area}: background \textit{before} any
    glitch rejection scheme is applied. \textbf{Black area}: \textit{after} the
    ``null-stream'' glitch rejection (see Sec.~\ref{sec:glitch}). \textbf{Gray
      area}: \textit{after} data-quality flags (see Sec.~\ref{sec:DQ}). CAT2 and
    3 refers to the different categories of the data-quality flags whose
    description goes beyond the scope of this article. \textbf{Bold curve}:
    expectation if noise is stationary and
    Gaussian. Credits:~\cite{aasi12:_virgo}.}
  \label{fig:background_burst}
\end{figure}

Besides the gravitational-wave channel $h(t)$, hundreds of auxiliary channels
including microphones, seismometers, magnetometers, etc. are recorded at any
given time during science data takings. Those channels can be used to get an
image of the operational and environmental status of the detector.  The observed
correlation between the GW channel with the auxiliary channels can help
determine the origin of noise artifacts and how the original disturbance couples
into the detector \cite{aasi12:_virgo}. A significant number of noise sources
are identified \textit{a posteriori} after the science data taking is
done. Those noise sources cannot be mitigated by fixing the
instrument. Instead, this leads to the development of a data-quality flag which,
when ``raised'', indicates that the data are improper, and any event occurring at
that time should be \textit{vetoed}.  This provides also an important resource
for background glitch rejection. Data-quality flags with a large ($\gg 1$)
efficiency (percentage of glitches vetoed by the flag) over dead-time (fraction
of science time rejected by the flag) ratio  are of particular interest
\cite{aasi12:_virgo}. About 200 data-quality flags are used in GW burst
searches. Figure~\ref{fig:background_burst} shows the background improvement
after vetoing.

\subsection{Results}
\label{sec:results}

\begin{figure}
  \centerline{\includegraphics[width=\textwidth]{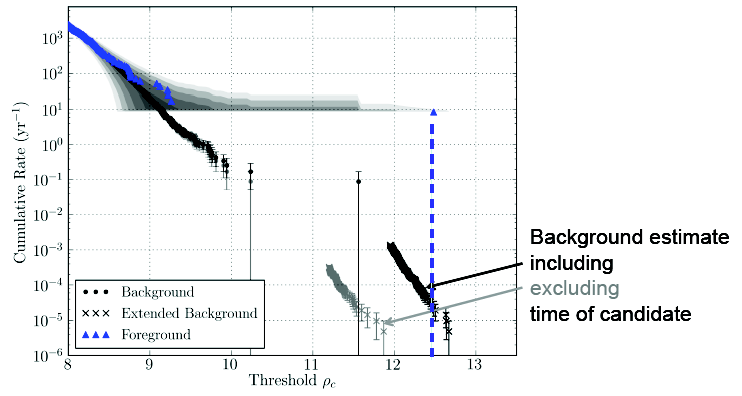}}
  \caption{\textbf{Cumulative event rate associated to the search for compact
      binary coalescences.} The ``candidate'' at $\rho_c\approx 12.5$ is a
    simulated signal inserted as part of a ``blind injection challenge
    exercise'' (see text). Credits: \cite{abadie12:_searc_ligos_virgos}}
  \label{fig:background_cbc}
\end{figure}

LIGO and Virgo conducted two joint science runs, labelled S5 for LIGO and VSR1
for Virgo for the first run, S6 and VSR2/3 for the second. A total of $T=635$
days of observing time have been
analyzed~\cite{abadie10:_all_ligo_geo_virgo,abadie12:_all_ligo_virgo}.  No GW
detection has been claimed yet. Upper limits (at 50\% confidence level) on the
GW strain obtained from an all-sky all-time GW burst search have been set. It is
slightly below $h_{rss} < 5 \times 10^{-22}$ Hz$^{-1/2}$ for frequency at about
200 Hz, where the bound is on the root-sum-square (rss) amplitude $h^2_{rss} \equiv
\int dt\: h^2_+(t) + h^2_\times(t)$ of the two GW polarizations at Earth. While
the exact result depends on the assumed GW model (here, a generic sine Gaussian
waveform characterized by its central frequency), it remains comparable for
waveforms with a similar spectrum.

This upper limit placed on a local quantity ($h_{rss}$ at Earth) can be
translated into astrophysical constraints: for instance, upper limits on the
radiated energy $E_{GW}$ by generic sources of linearly polarized GW located at
distance $d$. Averaging over the source inclination, the above strain limit
corresponds to $E_{GW}=2 \times 10^{-8} M_{\odot} c^2$ for galactic sources at
distance $d=10$ kpc, and $E_{GW}=5 \times 10^{-2} M_{\odot} c^2$ for source in
the Virgo cluster with $d=15$ Mpc. Those estimates are comparable to the
expected GW-radiated energy from core collapses and mergers of stellar-mass
compact objects respectively.

The same data, when searched specifically for inspiralling binaries of neutron
stars, leads to an upper limit on the rate of such astrophysical events of
${\cal R}_{90\%}=1.3 \times 10^{-4} \mathrm{yr}^{-1} \mathrm{Mpc}^{-3}$
\cite{abadie10:_searc_ligo_virgo_s5_vsr1,abadie12:_searc_ligos_virgos} which is
still two orders of magnitude larger than the rate estimate obtained from
population models \cite{abadie10:_predic}. Fig.~\ref{fig:background_cbc} shows
the cumulative rate of events detected by the matched-filtering procedure
outlined in Sec.~\ref{sec:timeseries} in coincidence in the H1 and L1 detectors
during four months of S6/VSR2-3 data taking. This distribution is displayed as a
function of the ranking statistic $\rho_c$ which combines the signal-to-noise
ratios measured at both detectors. The distribution of candidate events
(triangle) is superimposed to a background estimate (black dots) with error bars
(in gray). The last triangle on the right-hand side of the plot at $\rho_c
\approx 12.5$ is a candidate event detected with a false alarm rate of 1 in 7,000
years \cite{abadie12:_searc_ligos_virgos}. It was known in advance that a small
number of fake GW signals might be added ``blindly'' to the data. The exact time
and characteristics of these signals were only known by a small group of people
sworn to silence until the eventual ``opening of the envelope''. The envelope
was \textit{not} empty and contained the detected event code-named GW100916
\cite{11:_gw100916}.  Thanks to this exercise, it was possible to test the
entire decision-making chain up through the preparation of a publication.

\section{Online analysis and rapid electromagnetic follow-up}
\label{sec:emfollowup}

Sources of GW are likely sources of other kinds of emissions, such as
electromagnetic waves or jets of high-energy particles. The possible connection
between compact binary coalescences and GRB is an example
\cite{corsi11:_gravit}. This motivates cross-correlating GW with other types of
observations in the electromagnetic or neutrino spectra see e.g.,
\cite{abadie12:_searc_ligo_virgo,adrian-martinez12:_first_searc_gravit_waves_high}
for recent results. We will briefly report here on rapid follow-up observations
seeking electromagnetic counterparts to GW candidate events. A low-latency
analysis pipeline was operated for the first time during the last data taking
\cite{abadie12:_implem}. It allows to generate alerts ``on the fly'' within 20
minutes of the associated GW candidate event. Major changes were required with
respect to the original off-line pipelines. The most probable direction of the
source along with an error box were communicated to a dozen of partner
observatories \cite{abadie12:_implem} including radio telescopes, wide-field
optical telescopes and the X/gamma-ray satellite Swift. This has led to
follow-up observations which have been scanned for transient excursions. This
exercise has been extremely useful and will help to prepare the exciting future
of multi-messenger astronomy with the advanced detectors~\cite{12:_lsc_virgo,shawhan12:_rapid}.

\renewcommand{\small}{\fontsize{7}{10pt}\selectfont}

\begin{theacknowledgments} The authors gratefully acknowledge the
    support of the United States National Science Foundation for the
    construction and operation of the LIGO Laboratory, the Science and
    Technology Facilities Council of the United Kingdom, the Max-Planck-Society,
    and the State of Niedersachsen/Germany for support of the construction and
    operation of the GEO600 detector, and the Italian Istituto Nazionale di
    Fisica Nucleare and the French Centre National de la Recherche Scientifique
    for the construction and operation of the Virgo detector. The authors also
    gratefully acknowledge the support of the research by these agencies and by
    the Australian Research Council, the International Science Linkages program
    of the Commonwealth of Australia, the Council of Scientific and Industrial
    Research of India, the Istituto Nazionale di Fisica Nucleare of Italy, the
    Spanish Ministerio de Econom\'ia y Competitividad, the Conselleria
    d'Economia Hisenda i Innovaci\'o of the Govern de les Illes Balears, the
    Foundation for Fundamental Research on Matter supported by the Netherlands
    Organisation for Scientific Research, the Polish Ministry of Science and
    Higher Education, the FOCUS Programme of Foundation for Polish Science, the
    Royal Society, the Scottish Funding Council, the Scottish Universities
    Physics Alliance, The National Aeronautics and Space Administration, the
    National Research Foundation of Korea, Industry Canada and the Province of
    Ontario through the Ministry of Economic Development and Innovation, the
    National Science and Engineering Research Council Canada, the Carnegie
    Trust, the Leverhulme Trust, the David and Lucile Packard Foundation, the
    Research Corporation, and the Alfred P. Sloan Foundation. This paper has
    been assigned LIGO Document Number LIGO-P1200135 and Virgo TDS number
    VIR-0379C-12.
\end{theacknowledgments}

\bibliographystyle{unsrt}   

\makeatletter
\def\url@foostyle{%
  \@ifundefined{selectfont}{\def\UrlFont{\sf}}{\def\UrlFont{\footnotesize\ttfamily}}}
\makeatother

\urlstyle{foo}

\bibliography{paper}

\end{document}